\documentclass[a4paper,11pt,final,5p,times,twocolumn]{elsarticle}

\usepackage[T1]{fontenc}					
\usepackage{graphicx}
\usepackage{amsmath,amssymb,amsfonts}	
\usepackage{mathtools,physics} 				
\usepackage{subcaption} 			
\usepackage{booktabs}     
\usepackage{hyperref}
\usepackage{cleveref}						
\usepackage{orcidlink}						
\usepackage{bm,bbm}

\hypersetup{
colorlinks=true,
pdfauthor={Riasat Sheikh, Takashi Toma, Koji Tsumura},
pdftitle={Exothermic nuclear recoils from thermal quasi-pNGB dark matter}
}
\crefname{figure}{Fig.}{Figs.}
\Crefname{figure}{Figure}{Figures}
\crefname{table}{Table}{Tables}
\Crefname{table}{Table}{Tables}
\crefname{equation}{Eq.}{Eqs.}
\Crefname{equation}{Equation}{Equations}
\graphicspath{{./figs/}}
\biboptions{sort&compress}

\journal{Physics Letters B}
\begin{document}

\begin{frontmatter}
  \title{Exothermic nuclear recoils from thermal quasi-pNGB dark matter}

  \author[a]{Riasat Sheikh\corref{cor1}\,\orcidlink{0009-0007-1207-1358}}
  \ead{riasat.sheikh@phys.kyushu-u.ac.jp}
  \cortext[cor1]{Corresponding author.}

  \author[b,c]{Takashi Toma\,\orcidlink{0000-0001-5828-0090}}
  \ead{toma@staff.kanazawa-u.ac.jp}

  \author[a,d,e,f]{Koji Tsumura\,\orcidlink{0000-0003-3765-2750}}
  \ead{tsumura.koji@phys.kyushu-u.ac.jp}

  \affiliation[a]{organization={Department of Physics, Kyushu University},
    addressline={744 Motooka, Nishi-ku},
    city={Fukuoka},
    postcode={819-0395},
    country={Japan}}

  \affiliation[b]{organization={Institute of Liberal Arts and Science, Kanazawa University},
    city={Kanazawa},
    postcode={920-1192},
    country={Japan}}

  \affiliation[c]{organization={Institute for Theoretical Physics, Kanazawa University},
    city={Kanazawa},
    postcode={920-1192},
    country={Japan}}

  \affiliation[d]{organization={Research Center for Advanced Particle Physics, Kyushu University},
    addressline={744 Motooka, Nishi-ku},
    city={Fukuoka},
    postcode={819-0395},
    country={Japan}}

  \affiliation[e]{organization={Quantum and Spacetime Research Institute, Kyushu University},
    addressline={744 Motooka, Nishi-ku},
    city={Fukuoka},
    postcode={819-0395},
    country={Japan}}

  \affiliation[f]{organization={Kavli IPMU (WPI), UTIAS, University of Tokyo},
    city={Kashiwa},
    postcode={277-8584},
    state={Chiba},
    country={Japan}}

  \begin{abstract}
    We study a two-component dark matter scenario
    comprising a stable pseudo-Nambu-Goldstone boson (pNGB) and its
    nearly degenerate long-lived quasi-pNGB partner.
    An approximate scalar symmetry relates their masses and suppresses
    the tree-level Higgs-mediated elastic scattering of the excited
    state through the small mass splitting. Their coupled thermal
    evolution yields comparable relic populations, while requiring
    the observed total abundance fixes the dark scalar symmetry-breaking
    scale and hence the dark gauge coupling. At the studied benchmark,
    an excited fraction of approximately one half survives under the adopted
    evolution assumptions. The quasi-pNGB excited state can therefore produce
    exothermic nuclear recoils despite a suppressed tree-level elastic
    signal. As an application, we examine the high-energy LUX-ZEPLIN
    event. For relic compatible parameters, the recoil normalization fixes
    the kinetic mixing, allowing this interpretation to be tested by
    independent mediator searches.
  \end{abstract}

  \begin{keyword}
    Specific BSM Phenomenology \sep Dark Matter Direct Detection
    \par\noindent\textit{Preprint: } KYUSHU-HET-375 \sep KANAZAWA-26-08
  \end{keyword}

\end{frontmatter}

\section{Introduction}
\label{sec:intro}

An excited dark matter (DM) state can release its internal energy in a
nuclear recoil even when elastic scattering is suppressed
\cite{Tucker-Smith:2001myb, Finkbeiner:2009mi, Graham:2010ca}. In such
exothermic scattering, the transition to a lighter DM state supplies
energy in addition to the incident kinetic energy. The recoil spectrum
consequently depends on both the DM mass and the internal mass splitting,
opening a route to high-energy nuclear recoils. This mechanism requires an
excited population in the present halo.

A thermal realization must account for both the total abundance and the
relative populations of the two states. Annihilation removes DM particles,
whereas pair conversion redistributes them between the ground and excited
states; decay can further reduce the excited population. These processes
need not have the same efficiency. The question is therefore whether a
setup can produce the observed thermal abundance, retain enough excited
particles for exothermic scattering and suppress the accompanying elastic
signal. We address this in a two-component scalar dark sector whose near
degeneracy and tree-level elastic suppression have a common symmetry
origin.

The pseudo-Nambu-Goldstone boson (pNGB) class of DM provides a symmetry
origin for the cancellation of tree-level Higgs-mediated elastic
scattering at zero momentum transfer \cite{Gross:2017dan}. Loop-induced
scattering has been studied in the original model
\cite{Azevedo:2018exj,Ishiwata:2018sdi,Glaus:2020ihj} and related models
\cite{Abe:2022mlc,Abe:2024vxz,Glaus:2019itb,Glaus:2022rdc,Biekotter:2022bxp}.
A two-component pNGB DM scenario has also been studied in a softly broken
global $\mathrm{SU}(3)$ framework motivated by boosted dark matter
\cite{Sheikh:2026tjy}. The model proposed by Abe, Hamada and Tsumura (AHT)
\cite{Abe:2024vxz} provides the scalar and gauge interactions used here.
We study its nearly degenerate regime, containing the pNGB $a$ and the
quasi-pNGB excited partner $s_-$. Here we connect the thermal abundance,
population evolution and recoil signal quantitatively. Near degeneracy
suppresses the elastic scattering of $s_-$ while allowing comparable
thermal populations. The relic abundance fixes the dark scalar
symmetry-breaking scale and hence the dark gauge coupling, linking the
thermal solution to the strength of the inelastic recoil signal. The
coupled thermal history and the subsequent conversion and decay leave a
substantial surviving excited population at the studied benchmark.

As an application, we consider the recently reported LUX-ZEPLIN (LZ) event
at $E_R = 248 \pm 23_{\rm stat} \pm 23_{\rm sys}~\mathrm{keV}$ in an
exposure of $2.84~\mathrm{tonne\,yr}$ with $2.6\sigma$ global
significance. The LZ analysis considers non-relativistic effective
momentum-suppressed interactions that can enhance the relative
contribution of high-energy recoils \cite{LZ:2026axp}. Other
interpretations of this signal study spectra \cite{Su:2026rwz,
Dent:2026bji}, exothermic fermions \cite{deLima:2026shq, Baer:2026fpy,
Gemmell:2026yaw}, scalar dark sectors \cite{Das:2026buc, Zhu:2026dag}, and
population evolution \cite{An:2026pkc, Oliveira:2026jtn}. Solar capture
and neutrino observations also constrain inelastic Higgsino
interpretations \cite{Pospelov:2026ewn}. Using an approximate
detector-response model and accounting for the event date and empty
sideband \cite{Dent:2026bji, McCabe:2026crm}, we examine whether the
surviving excited population can account for this event. The required
kinetic mixing connects this interpretation to independent mediator
searches.

\section{Model}
\label{sec:model}

The AHT model contains the SM Higgs doublet $H$ and two SM-singlet complex
scalars $\phi_{1,2}^{}$ with equal unit dark gauge $\mathrm{U}(1)_V$
charges, opposite global $\mathrm{U}(1)$ charges and an exchange symmetry
$\phi_1^{} \leftrightarrow \phi_2^{}$ \cite{Abe:2024vxz}. The scalar
representations and charges are listed in \cref{tab:charges}.

\begin{table}[t]
  \centering
  \small
  \setlength{\tabcolsep}{3pt}
  \renewcommand{\arraystretch}{1.2}
  \begin{tabular}{@{}lccccc@{}}
    \toprule
    & $\mathrm{SU}(3)_c$ & $\mathrm{SU}(2)_L$ & $\mathrm{U}(1)_Y$
    & $\mathrm{U}(1)^{\rm global}$ & $\mathrm{U}(1)_V^{\rm gauge}$     \\
    \midrule
    $H$      & $\mathbf{1}$ & $\mathbf{2}$ & $+\tfrac12$ & $0$  & $0$  \\
    $\phi_1$ & $\mathbf{1}$ & $\mathbf{1}$ & $0$         & $+1$ & $+1$ \\
    $\phi_2$ & $\mathbf{1}$ & $\mathbf{1}$ & $0$         & $-1$ & $+1$ \\
    \bottomrule
  \end{tabular}
  \caption{Representations and charges of the scalar fields.}
  \label{tab:charges}
\end{table}

The scalar potential is given by
\begin{equation}
  \begin{aligned}
    \mathcal V ={} & -m_H^2\abs{H}^2 + \lambda_H\abs{H}^4
    + m_1^2\qty(\abs{\phi_1}^2 + \abs{\phi_2}^2)                                \\
                   & + \frac{\lambda_1}{2}\qty(\abs{\phi_1}^4 + \abs{\phi_2}^4)
    + \lambda_3\abs{\phi_1}^2\abs{\phi_2}^2                                     \\
                   & + \underbrace{~\lambda_{H1}\abs{H}^2
    \qty(\abs{\phi_1}^2 + \abs{\phi_2}^2)}_\text{Higgs portal}
    ~-\underbrace{~m_{12}^2\qty(\phi_1^*\phi_2^{} + \mathrm{h.c.})~}_{\text{soft term}}\,.
  \end{aligned}
  \label{eq:scalar-potential-main}
\end{equation}
Here $m_H^2$ and $m_1^2$ are quadratic parameters,
$\lambda_H$ is the Higgs quartic, $\lambda_{H1}$ is the Higgs portal,
and $\lambda_1$ and $\lambda_3$ are the singlet self- and
mixed quartic couplings. The soft parameter $m_{12}^2$ explicitly
breaks the global symmetry.

We take equal vacuum expectation values (VEVs) for the singlet scalars,
$\expval{\phi_{1,2}^{}} = \flatfrac{v_s^{}}{2}$, preserving the exchange
symmetry and breaking the dark gauge symmetry. In unitary gauge, the
scalar fields can be written as
\begin{equation}
  \begin{gathered}
    H = \frac{1}{\sqrt2}\,\mqty(0\\ v_h^{} + h) \,, \\
    \phi_1^{} = \frac{v_s^{} + s_+ + s_- + i a}{2}\,, \quad
    \phi_2^{} = \frac{v_s^{} + s_+ - s_- - i a}{2}\,.
  \end{gathered}
  \label{eq:scalar-fields-main}
\end{equation}

\begin{figure}
  \centering
  \includegraphics[width=\linewidth]{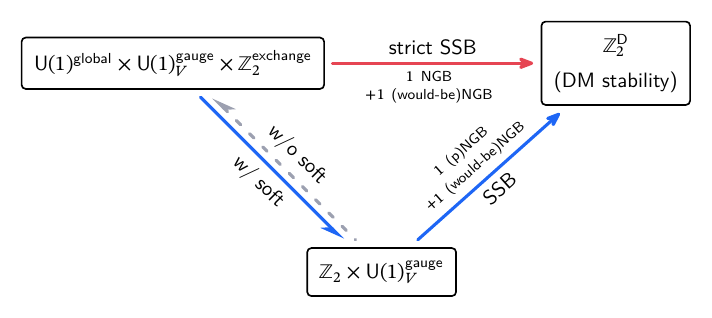}
  \caption{Summary of the symmetry-breaking pattern.}
  \label{fig:symmetry-breaking-flow}
\end{figure}

\Cref{fig:symmetry-breaking-flow} summarizes the symmetry-breaking
pattern of the AHT model. Without the soft term in
\cref{eq:scalar-potential-main}, the singlet VEVs produce one physical
Nambu-Goldstone boson (NGB)
and one would-be NGB. The common singlet phase is eaten by the dark gauge
boson, while the relative phase $a$ is the physical mode. The soft term
explicitly breaks the global symmetry and gives $a$ its pNGB mass. The
radial excitation $s_-$ and $a$ are both odd under the unbroken $\mathbb{Z}_2^{\rm D}$
symmetry, whose dark parity ensures the stability of the lighter state
$a$. The longevity of $s_-$ additionally depends on its available decay
channels. Their masses are
\begin{equation}
  m_a^2 = 2m_{12}^2\,,\qquad
  m_{s_-}^2 = 2m_{12}^2 + \frac{1}{2}\qty(\lambda_1 - \lambda_3)v_s^2\,.
  \label{eq:quasi-pngb-masses}
\end{equation}

The even singlet $s_+$ mixes with the Higgs fluctuation $h$ through the
portal coupling and the mass eigenstates are defined by
\begin{equation}
  \mqty(h_1\\ h_2)
  = \mqty(\cos\theta & \sin\theta\\ -\sin\theta & \cos\theta)
  \mqty(h\\ s_+),
  \label{eq:scalar-mixing-main}
\end{equation}
where $\theta$ is the mixing angle.
Their masses are $m_{h_1}^{}$ and $m_{h_2}^{}$, with $h_1^{}$ identified as
the observed SM Higgs, $m_{h_1}^{} = 125~\mathrm{GeV}$ and
$v_h^{} = 1/\sqrt{\mathstrut\sqrt2 G_F^{}}\approx246~\mathrm{GeV}$,
where $G_F^{}$ is the Fermi constant.

Kinetic mixing $\sin\epsilon$ between hypercharge and the dark gauge field
couples the dark gauge boson $Z'$ to the SM. At small mixing and
$m_{Z'}^{} \ll m_Z^{}$, the dark gauge coupling is $g_V^{} \simeq
\flatfrac{m_{Z'}^{}}{v_s^{}}$ and the coherent vector coupling to a
nucleus is $\mathcal C_N \simeq - Z e \varepsilon_\gamma$, where
$\varepsilon_\gamma = \cos\theta_W \sin\epsilon$, $Z$ is the atomic
number, $e$ is the electromagnetic coupling and $\theta_W$ is the weak
mixing angle. The derivative gauge interaction $\mathcal L \supset g_V^{}
Z'_\mu \qty(s_- \partial^\mu a - a \partial^\mu s_-)$ permits both
endothermic and exothermic scattering.

With $m_{h_1}^{}$ and $v_h^{}$ fixed to their SM values, the independent
model inputs in the degenerate limit are
\begin{equation}
  \begin{gathered}
    m_a^{},\quad v_s^{},\quad m_{h_2}^{}, \quad
    m_{Z'}^{},\quad \sin\theta,\quad \sin\epsilon \,.
  \end{gathered}
\end{equation}

In the limit $\lambda_1 = \lambda_3$, the enhanced global $\mathrm{O}(4)$
symmetry makes $a$ and $s_-$ the two degenerate real components of a
complex pNGB, as discussed in \cite{Abe:2024vxz}. A small departure from
this generates the splitting
\begin{equation}
  \Delta m \equiv m_{s_-}^{} - m_a^{} > 0\,,
  \label{eq:splitting-definition}
\end{equation}
which we use as an additional independent parameter.
We therefore refer to $s_-$ as a quasi-pNGB.

\section{Relic abundance and excited-state survival}
\label{sec:relic-abundance}

We use \texttt{FeynRules} \cite{Alloul:2013bka} to generate the model
files and solve the separate Boltzmann equations for $a$ and $s_-$ with
\texttt{micrOMEGAs} \cite{Alguero:2023zol}, including annihilation,
coannihilation and conversion and assuming a common bath temperature. We
fix
\begin{equation}
  \begin{gathered}
    m_{h_2}^{} = 300~\mathrm{GeV},\quad m_{Z'}^{} = 0.27~\mathrm{GeV},\\
    \sin\theta = 0.1,\quad \sin\epsilon = 10^{-5},
  \end{gathered}
  \label{eq:relic-benchmark-inputs}
\end{equation}
and vary $v_s^{}$ to obtain the observed relic abundance $\Omega_{\rm tot}h_{\rm hub}^2
  = 0.120 \pm 0.001$ \cite{Planck:2018vyg}. Here $h_{\rm hub}$ is the reduced Hubble
constant, $\Omega_i$ are the density parameters and
$\Omega_{\rm tot} = \Omega_a + \Omega_{s_-}$.
The $m_{Z'}^{}$ choice is motivated by the momentum transfer
$\abs{\vb q}=\sqrt{\mathstrut 2 m_N^{} E_R^{}} \simeq 0.22$--$0.27~\mathrm{GeV}$ for
true xenon recoils of $200$--$300~\mathrm{keV}$, where $m_N^{}$ is the nuclear
mass and $E_R$ is the recoil energy. At fixed $v_s^{}$ and
mixing, the scattering factor
$m_{Z'}^2 \, v_s^{-2} \qty(\abs{\vb q}^2 + m_{Z'}^2)^{-2}$ is largest at
$m_{Z'}^{} = \abs{\vb q}$.

\begin{figure}[h]
  \centering
  \includegraphics[width=\linewidth]{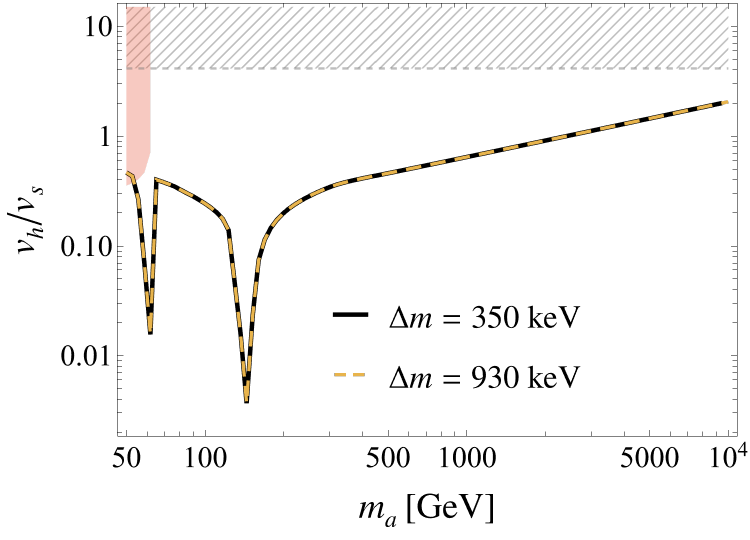}
  \caption{relic compatible $v_h^{}/v_s^{}$
  for $\Delta m = 350~\mathrm{keV}$ (black solid) and
  $930~\mathrm{keV}$ (yellow dashed).
  Red shading and gray hatching mark the regions excluded by
  Higgs-invisible decays and perturbative unitarity, respectively.}
  \label{fig:parameter-scan}
\end{figure}

\Cref{fig:parameter-scan} shows the ratio $v_h^{}/v_s^{}$ required
by the total relic abundance. The black solid and yellow dashed curves
represent the $\Delta m = 350~\mathrm{keV}$ and $930~\mathrm{keV}$ relic solutions,
respectively. The two curves nearly coincide because both splittings satisfy
$\Delta m\ll T_f$, where $T_f$ is the freeze-out temperature. The relative equilibrium
abundance is therefore barely affected by Boltzmann suppression, making
the relic solution insensitive to the change in $\Delta m$. Near
$m_a^{}=m_{h_{1,2}}^{} / 2$, resonant annihilation permits a larger
$v_s^{}$, producing the two dips. The red region is excluded by
Higgs-invisible decays, while the gray hatching shows the perturbative
unitarity limit on the scalar quartic and dark gauge couplings.
Since $m_{Z'}^{} \ll m_a^{}$, the annihilation channels
$aa, s_-s_- \to Z'Z'$ are open throughout the scanned mass range.
The relic abundance determines $v_s^{}$ and hence $g_V^{} \simeq
  m_{Z'}^{}/v_s^{}$. This permits a small gauge coupling while maintaining
efficient annihilation into longitudinal $Z'$ states, whose interactions
depend on $g_V^{}/m_{Z'}^{} \simeq 1/v_s^{}$.

The mass-weighted relic fractions are given by
\begin{equation}
  \xi_i^{} = \frac{\Omega_i}{\Omega_{\rm tot}}
  =\frac{m_i^{} Y_i}{m_a^{} Y_a + m_{s_-}^{} Y_{s_-}}.
  \label{eq:relic-fraction}
\end{equation}
Here $i \in \qty{a,s_-}$ and $Y_i = n_i/\mathfrak{s}(T)$ are the comoving
yields, with number densities $n_i$ and entropy density $\mathfrak{s}(T)$
at bath temperature $T$. The fractions satisfy $\xi_a + \xi_{s_-} = 1$.
The coupled scan gives $\xi_a \simeq \xi_{s_-} \simeq 1/2$, consistent with
nearly equal equilibrium weights for $\Delta m\ll T\ll m_a$.

\begin{figure}[h]
  \centering
  \includegraphics[width=\linewidth]{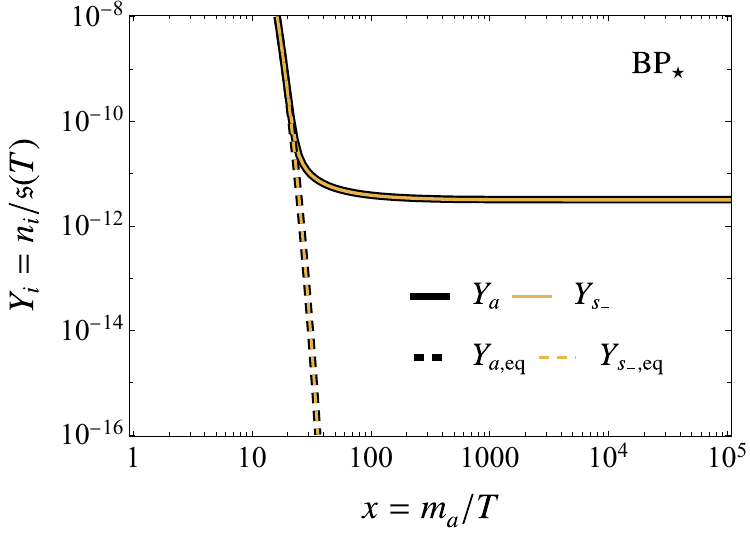}
  \caption{Coupled thermal yields at $\mathrm{BP}_\star$.
  Black (yellow) denotes yields for $a$ ($s_-$) and solid (dashed) curves show
  calculated (equilibrium) yields.}
  \label{fig:parameter-scan-b}
\end{figure}

\Cref{fig:parameter-scan-b} shows freeze-out at
$\mathrm{BP}_\star$ with $m_a^{} = 70~\mathrm{GeV}$, $\Delta m =
  930~\mathrm{keV}$, and relic compatible $v_s^{} \simeq 651~\mathrm{GeV}$.
As $x = m_a^{}/T$ increases, the calculated yields depart from their
rapidly falling equilibrium values and approach constant relic yields. The
yields of $a$ (black) and $s_-$ (yellow) nearly overlap throughout this
evolution and both species freeze out with comparable populations, giving
an excited-state relic fraction $\xi_{s_-} \simeq 0.498$. Efficient
annihilation therefore coexists with a substantial excited population.

\begin{figure}[h]
  \centering
  \includegraphics[width=\linewidth]{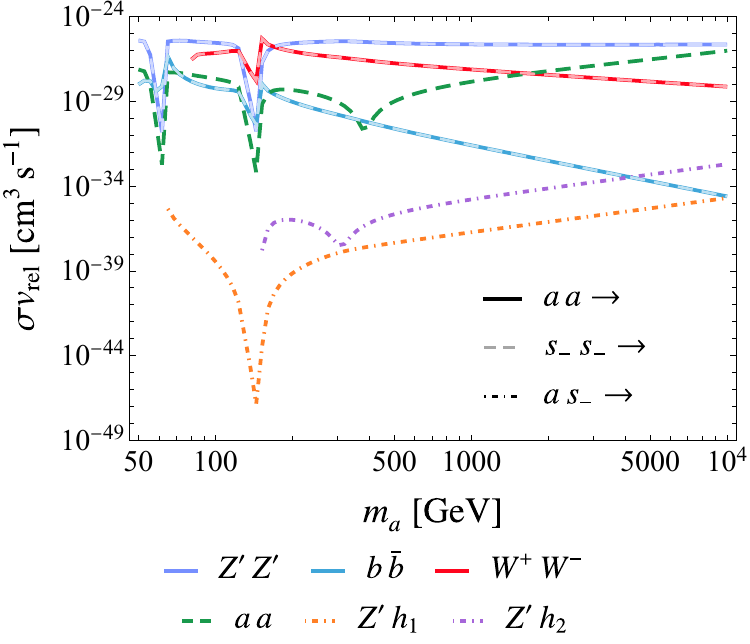}
  \caption{Present-day $\sigma v_{\rm rel}$ along the
  $\Delta m = 930~\mathrm{keV}$ relic solution.
  For the common annihilation final states, solid (dashed) curves denote
  incoming $aa$ ($s_-s_-$). Green dashed shows pair conversion; orange and purple
  dot-dashed curves show coannihilation.}
  \label{fig:XSec}
\end{figure}

\Cref{fig:XSec} shows the annihilation, conversion and coannihilation
cross sections at present-day halo velocities ($v_{\rm rel} \simeq 10^{-3}c$)
along the relic solution for $\Delta m = 930~\mathrm{keV}$.
The conversion and coannihilation rates illustrate the interference
structure. At zero incoming velocity and for $\Delta m\ll m_{Z'}^{}\ll
  m_a^{}$, the tree-level rates away from resonances and final-state
thresholds are approximately given by
\begin{align}
  (\sigma v_{\rm rel})_{s_-s_- \to aa}
   & \simeq \frac{m_a^2}{4\pi v_s^4}\sqrt{\frac{2\Delta m}{m_a^{}}}
  \nonumber                                                         \\
   & \quad\times\qty[\frac{2 m_{Z'}^2}{m_{Z'}^2 + 2 m_a^{}\Delta m}
  -\sum_{i = 1,2}\frac{\kappa_i^2\, m_{h_i}^2}{4 m_a^2 - m_{h_i}^2}]^2,
  \label{eq:conversion-rate-main}                                   \\
  (\sigma v_{\rm rel})_{as_- \to Z'h_i}
   & \simeq \frac{\kappa_i^2\,(\Delta m)^2}{128\pi m_a^4 v_s^4}
  \qty(1 - \frac{m_{h_i}^2}{4 m_a^2})
  \nonumber                                                         \\
   & \quad\times\qty[\frac{m_{h_i}^4+16m_a^2m_{h_i}^2-16m_a^4}
  {4m_a^2-m_{h_i}^2}]^2,
  \label{eq:coannihilation-rate-main}
\end{align}
where $\kappa_1^{} = \sin\theta$ and $\kappa_2^{} = \cos\theta$.
The coannihilation expression applies to open channels and retains
longitudinal $Z'$ production. Finite-velocity contributions prevent the
plotted rates from vanishing at the zero-velocity cancellations.

The blue, light-blue and red curves in \cref{fig:XSec} show annihilation
into $Z'Z'$, $b\bar b$ and $W^+ W^-$, respectively. The nearly overlapping
solid ($aa$) and dashed ($s_-s_-$) curves indicate similar annihilation
rates for the two species. For annihilation into longitudinal $Z'$ pairs,
the leading contact and crossed-scalar contributions cancel, leaving an
amplitude controlled by Higgs exchange. The green (dashed) conversion
channel $s_-s_- \to aa$ has restricted phase space, reflected in the
prefactor $\sqrt{\mathstrut 2\Delta m/m_a^{}}$ of
\cref{eq:conversion-rate-main}. Similarly, the orange $as_-\to Z'h_1$ and
purple $as_-\to Z'h_2$ dot-dashed coannihilation channels exhibit the
$(\Delta m)^2$ suppression in the zero-velocity approximation of
\cref{eq:coannihilation-rate-main}.

At $\mathrm{BP}_\star$, pair annihilation into $Z'Z'$ gives $\sigma v_{\rm
rel}\sim10^{-26}~\mathrm{cm^3\,s^{-1}}$ for both species. Conversion and
the open coannihilation channel are respectively $\mathcal O(10^{-2})$ and
$\mathcal O(10^{-11})$ relative to this rate at halo velocities. Across
the mass scan, cancellation in the bracket of
\cref{eq:conversion-rate-main} explains the additional green minimum near
$m_a^{} \simeq 375~\mathrm{GeV}$, where the crossed $Z'$ exchange
interferes destructively with the combined scalar contact and $s$-channel
Higgs contributions. For coannihilation, the numerator in
\cref{eq:coannihilation-rate-main} vanishes near $m_a^{} \simeq
1.029\,m_{h_i}^{}$, corresponding to $129$ and $309~\mathrm{GeV}$ for
$h_1$ and $h_2$. These cancellations arise between $s$-channel vector
exchange and crossed scalar exchange. The dip of the orange curve also
reflects the increased relic compatible $v_s^{}$ near the heavier-Higgs
resonance, which reduces the coannihilation rate through its $v_s^{-4}$
dependence.

\paragraph{Elastic scattering of $a$ and $s_-$}
The pNGB cancellation suppresses tree-level elastic scattering of $a$
\cite{Gross:2017dan}, and for the quasi-pNGB $s_-$, the zero-momentum
amplitude is proportional to $m_{s_-}^2 - m_a^2$. At zero momentum
transfer, this amplitude can be matched to a real-singlet Higgs portal
\cite{Cline:2013gha}, $\lambda_{HS}^{\rm eff}\abs{H}^2s_-^2$, with
\begin{equation}
  \lambda_{HS}^{\rm eff}
  \simeq \qty(\cfrac{v_{h}^{}}{v_s^{}})^{}
  \qty(1-\frac{m_{h_1}^2}{m_{h_2}^2})
  \cfrac{m_a^{} \Delta m }{v_h^2}
  \sin 2\theta,
  \label{eq:effective-elastic-portal}
\end{equation}
for $\Delta m\ll m_a$.
At fixed $m_a^{}$, Higgs masses, mixing and $v_s^{}$, the effective
coupling is linear in $\Delta m$, so the tree-level zero-momentum elastic
cross section scales as $(\Delta m)^2$.

At $\mathrm{BP}_\star$, the relic compatible solution gives a tree-level
Higgs-mediated elastic cross section at zero momentum transfer of
$\sigma_{\rm SI}^{s_-} \sim 10^{-56}~\mathrm{cm^2}$, far below the LZ
limit even when the full local DM density is assigned to $s_-$
\cite{LZ:2024zvo}.

The excited state must also survive against decay. The two-body decay
$s_-\to aZ'$ requires $\Delta m>m_{Z'}^{}$ and is closed for the
benchmarks considered here. The excited state can still decay through an
off-shell $Z'$. For $\Delta m<2m_e^{}$, where $m_e^{}$ is the electron
mass, the electron-pair channel is also closed, while neutrino and
radiative decays are strongly suppressed
\cite{Finkbeiner:2009mi,deLima:2026shq}.

\section{Inelastic scattering}
\label{sec:detection-prospects}

The minimum halo speed for either scattering direction is
\begin{equation}
  v_\text{DM}^{\min,\,\pm}(E_R)
  =\cfrac{1}{\sqrt{\mathstrut 2 m_N^{} E_R}}\,
  \abs{\cfrac{m_N^{} E_R}{\mu_N}\pm\Delta m}.
  \label{eq:vmin-main}
\end{equation}
Here $v_{\rm DM}^{}$ is the incident DM speed,
$\mu_N^{}=m_a^{}m_N^{}/(m_a^{}+m_N^{})$ is the reduced mass to leading
order in the small splitting, and the $+$ ($-$) sign denotes endothermic scattering or
upscattering (exothermic or downscattering). Upscattering
requires $v_{\rm DM} \ge \sqrt{\mathstrut 2 \Delta m/\mu_N^{}}$ and
downscattering has zero minimum speed at $E_R=\mu_N\Delta m/m_N$.
At leading non-relativistic order, the $Z'$-mediated differential cross
section is
\begin{equation}
  \dv{\sigma_N^{}}{E_R}
  \simeq \frac{m_N^{}}{2\pi v_{\rm DM}^2}
  \qty(\cfrac{g_V^{} \mathcal C_N}{q^2 + m_{Z'}^2})^2
  \qty[F_{\rm H}^{(N)}(q)]^2,
  \label{eq:nr-cross-section}
\end{equation}
where $q^2 = 2 m_N^{} E_R$ and $F_{\rm H}^{(N)}$ is the Helm form factor.
At fixed masses and halo fractions, the differential recoil rate scales as
\begin{equation}
  \dv{R}{E_R} \propto \qty(g_V^{} \sin\epsilon)^2.
  \label{eq:coupling-product}
\end{equation}
The relic solution fixes $g_V^{}$, and fitting the signal determines the
mixing for the specified halo fractions.

We weight the rates of both species by their relic fractions and sum over
natural xenon isotopes using a truncated Standard Halo Model with local DM
density $\rho_{\rm DM}^{} = 0.3~\mathrm{GeV\,cm^{-3}}$, speed scale
$v_0^{} = 238~\mathrm{km\,s^{-1}}$ and Galactic escape speed $v_{\rm
esc}^{} = 544~\mathrm{km\,s^{-1}}$. Expected event counts are averaged
over the run from 27 March 2023 to 1 April 2024, assuming uniformly
distributed live time, while the event density is evaluated on 16 June
2023 \cite{McCabe:2026crm}. We fold the recoil spectrum with an
approximate Gaussian response $\sigma_E^{} = 1.46
\sqrt{E_R/\mathrm{keV}}~\mathrm{keV}$ and an approximate acceptance
defined as
\begin{equation}
  \begin{aligned}
    \varepsilon_{\rm LZ}^{}(E_R)
     & =\varepsilon_0^{}\,
    \frac12
    \operatorname{erfc}
    \qty[\frac{E_R-E_{\rm hi}}{\sqrt2\,\sigma_E^{}(E_{\rm hi})}] \\
     & \qquad \qquad
    \frac12\operatorname{erfc}
    \qty[\frac{E_{\rm lo}-E_R}{\sqrt2\,\sigma_E^{}(E_{\rm lo})}],
  \end{aligned}
  \label{eq:lz-efficiency-model}
\end{equation}
with $\varepsilon_0^{} = 0.96$, $E_{\rm lo} = 5.4~\mathrm{keV}$ and
$E_{\rm hi}=269.9~\mathrm{keV}$.
This reproduces the reported $96\%$ plateau and smooth roll-offs near the reported $50\%$ efficiency thresholds
\cite{LZ:2026axp}. The
event-energy density also includes the $23~\mathrm{keV}$ systematic uncertainty.

Following the signal-only treatment of \cite{Dent:2026bji}, our diagnostic
likelihood approximates the signal-like data by one event at
$248~\mathrm{keV}$ in the search selection and zero events in the
$350$--$680~\mathrm{keV}$ sideband, neglecting backgrounds
\cite{LZ:2026axp}. For the sideband signal efficiency we assume the $96\%$
plateau extends to this region \cite{Dent:2026bji}. The intervening
$270$--$350~\mathrm{keV}$ gap provides no additional counting constraint.
Let $N_{\rm win}$ be the expected count after search acceptance and
$N_{\rm SB}$ the expected reconstructed count in the sideband, both
averaged over the run, with $r_{\rm SB}^{} = N_{\rm SB}/N_{\rm win}$. The
extended one-event likelihood, up to a parameter-independent factor, is
\begin{equation}
  \begin{aligned}
    \mathcal L
     & = N_{\rm win}\, p_{248}^{}
    \exp\qty[-N_{\rm win}\qty(1 + r_{\rm SB}^{})], \\
    p_{248}^{}
     & \equiv \frac{1}{N_{\rm win}}
    \eval{\dv{N}{E_{\rm obs}}}_{E_{\rm obs} = 248~\mathrm{keV}}\,,
  \end{aligned}
  \label{eq:one-event-likelihood-main}
\end{equation}
where $\dv*{N}{E_{\rm obs}}$ is the exposure-scaled accepted event density
on 16 June 2023, including the statistical and systematic energy
uncertainties. Its normalization uses the run-averaged $N_{\rm win}$,
retaining the event-date dependence \cite{McCabe:2026crm}.
At fixed spectral shape, $p_{248}^{}$ and $r_{\rm SB}^{}$ are independent of
the overall normalization. Profiling it at each mass and splitting gives
\begin{equation}
  \begin{gathered}
    \widehat N_{\rm win} = \frac{1}{1 + r_{\rm SB}^{}},\qquad
    \widehat N_{\rm SB} = r_{\rm SB}^{} \widehat N_{\rm win}, \\
    \widehat{\mathcal L} = e^{-1}\frac{p_{248}^{}}{1 + r_{\rm SB}^{}},
  \end{gathered}
  \label{eq:profiled-likelihood}
\end{equation}
where hats denote quantities evaluated at the best-fitting normalization.
The factor $e^{-1}$ is common to all profiled scan points and
cancels in likelihood ratios. The likelihood therefore favors
spectral support at the observed event energy and penalizes
a large sideband-to-search yield ratio.

\section{Signal region and predictions}
\label{sec:signal-results}

\begin{table}[h]
  \centering
  \small
  \setlength{\tabcolsep}{5pt}
  \begin{tabular}{cccc}
    \toprule
    Parameter & $\mathrm{BP}_\star$ & $\mathrm{BP}_\square$
    & $\mathrm{BP}_\triangle$                                            \\
    \midrule
    $m_a^{}~[\mathrm{GeV}]$   & $70$    & $1000$   & $70$                \\
    $\Delta m~[\mathrm{keV}]$ & $930$   & $350$    & $200$               \\
    $v_s^{}~[\mathrm{GeV}]$   & $651$   & $381$    & $651$               \\
    $g_V^{}$ & $4.15\times10^{-4}$ & $7.09\times10^{-4}$
    & $4.15\times10^{-4}$                                                \\
    $\varepsilon_\gamma$ & $1.06\times10^{-5}$ & $6.34\times10^{-6}$
    & $6.21\times10^{-7}$                                                \\
    $\xi_a$                   & $0.502$ & $0.501$  & $0.500$             \\
    $\xi_{s_-}$               & $0.498$ & $0.499$  & $0.500$             \\
    $r_{\rm SB}^{}$           & $0.944$ & $0.0450$ & $9.30\times10^{-8}$ \\
    \bottomrule
  \end{tabular}
  \caption{Benchmark values at $m_{Z'}^{}=0.27~\mathrm{GeV}$,
  $m_{h_2}^{}=300~\mathrm{GeV}$ and $\sin\theta=0.1$.
  The relic solution determines $v_s^{}$, $g_V^{}$ and both relic fractions;
  $\varepsilon_\gamma$ follows from the fitted recoil normalization.}
  \label{tab:benchmarks}
\end{table}

We compare $\mathrm{BP}_\star = \qty(70~\mathrm{GeV},930~\mathrm{keV})$
with $\mathrm{BP}_\square^{} = \qty(1000~\mathrm{GeV},350~\mathrm{keV})$
and $\mathrm{BP}_\triangle = \qty(70~\mathrm{GeV},200~\mathrm{keV})$,
where each pair denotes $\qty(m_a^{},\Delta m)$. For each benchmark, we
determine $v_s^{}$ from the relic abundance and fit $\sin\epsilon$ using
the recoil-event likelihood thus determining the $\varepsilon_\gamma^{}$.
The benchmark parameters and sideband ratios are collected in
\cref{tab:benchmarks}.

\begin{figure}[h]
  \centering
  \includegraphics[width=\linewidth]{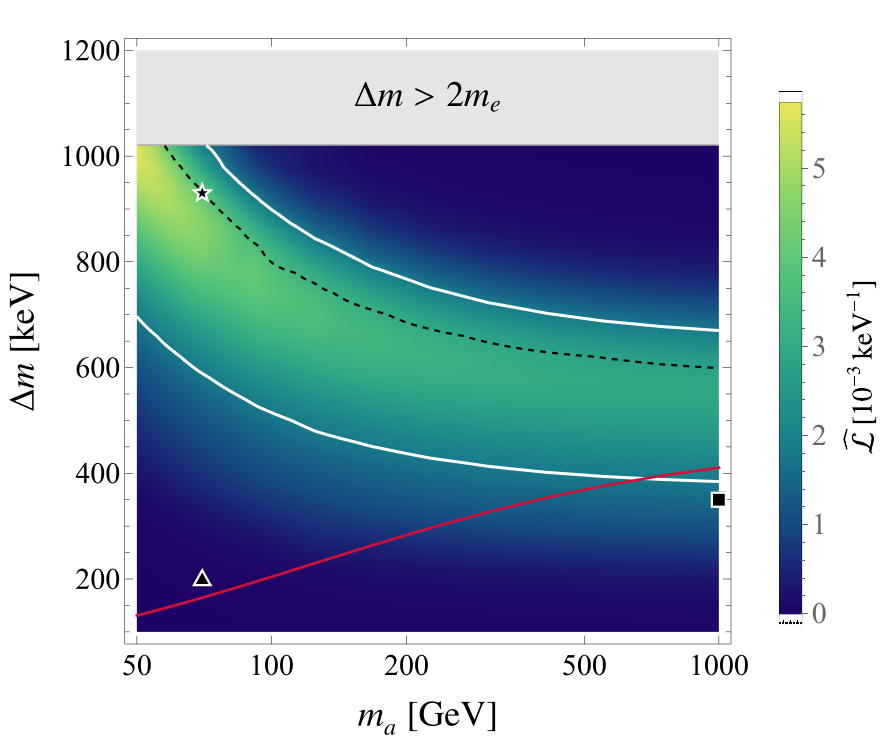}
  \caption{Relic compatible mass splitting likelihood.
  Color shows the profiled likelihood.
  Star, square and triangle mark the corresponding benchmarks in
  \cref{tab:benchmarks}.}
  \label{fig:signal-likelihood}
\end{figure}

\Cref{fig:signal-likelihood} shows the relic compatible profiled
likelihood, with brighter colors indicating greater spectral support
after the sideband penalty. The black $r_{\rm SB} = 1$ contour marks equal expected
search and sideband counts. Above the red boundary, only downscattering
is accessible for the adopted halo cutoff.
White contours indicate the
$-2\ln(\widehat{\mathcal L}/\widehat{\mathcal L}_{\max})=2.3$
relative likelihood to the scan maximum.
The gray strip marks $\Delta m > 2 m_e^{}$, where electron-pair decay opens.
$\mathrm{BP}_\star$ lies in the favored
low-mass, large-splitting band, where exothermic recoils
provide spectral support at the observed event energy
and a comparable expected sideband yield ($r_{\rm SB} \simeq 0.944$),
while retaining a substantial excited quasi-pNGB population and
suppressed tree-level elastic scattering.

\begin{figure}
  \centering
  \includegraphics[width=\linewidth]{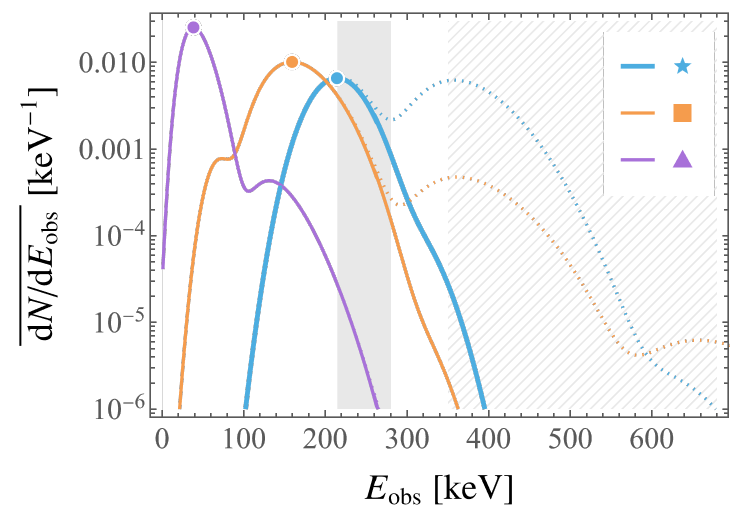}
  \caption{Reconstructed recoil spectra for
  $\mathrm{BP}_\star$ (blue), $\mathrm{BP}_\square^{}$ (orange) and
  $\mathrm{BP}_\triangle$ (purple) as in \cref{tab:benchmarks}. Solid (dotted) curves include
  (omit) the energy-dependent search-selection acceptance.
  Circles mark the solid-curve maxima. The gray band is $248 \pm
    32.5~\mathrm{keV}$, combining the statistical and systematic event
  uncertainties in quadrature. Hatching marks the
  $350$--$680~\mathrm{keV}$ sideband.}
  \label{fig:signal-spectra}
\end{figure}

In \cref{fig:signal-spectra} we show the reconstructed spectra for the
three benchmark points, $\mathrm{BP}_\star$ (blue),
$\mathrm{BP}_\square^{}$ (orange) and $\mathrm{BP}_\triangle$ (purple),
respectively. The displayed spectra are averaged over the run period,
assuming uniform live-time weighting average
\begin{equation}
  \overline{\dv{N}{E_{\rm obs}}}
  = \qty(t_2^{} - t_1^{})^{-1}
  \int_{t_1^{}}^{t_2^{}} \dd t\, \dv{N}{E_{\rm obs}}(E_{\rm obs},t)\,,
\end{equation}
where $t_1^{}$ and $t_2^{}$ correspond to 27 March 2023 and 1 April 2024,
respectively. The time-dependent density is normalized to the total
exposure. This average accounts for the changes in the predicted recoil spectrum during the run,
whereas the likelihood uses the spectrum evaluated on the event date, as described above.
The solid curves include the energy-dependent acceptance modeled in \cref{eq:lz-efficiency-model}.
The dotted curves instead use a constant acceptance of $96\%$,
equal to the plateau value assumed for the sideband.
The circular markers identify the maxima
of the accepted spectra. At fixed $m_a^{}=70~\mathrm{GeV}$, reducing the splitting from $930$ to
$200~\mathrm{keV}$ shifts the purple spectrum to lower energies and
greatly reduces its support near the observed event. At
$\mathrm{BP}_\star$, the accepted spectrum peaks at $E_{\rm obs} \simeq
  214.5~\mathrm{keV}$ and provides spectral support at the observed
$248~\mathrm{keV}$. The dotted blue curve also shows a higher-energy
component in the hatched sideband, as discussed earlier, with comparable fitted counts,
$\widehat N_{\rm win} \simeq 0.515$ and $\widehat N_{\rm SB} \simeq
  0.485$. A future search with a calibrated high-energy response could test
the spectral shape and relative yield of this higher-energy component
\cite{Dent:2026bji, deLima:2026shq}.

\begin{figure}[t]
  \centering
  \includegraphics[width=\linewidth]{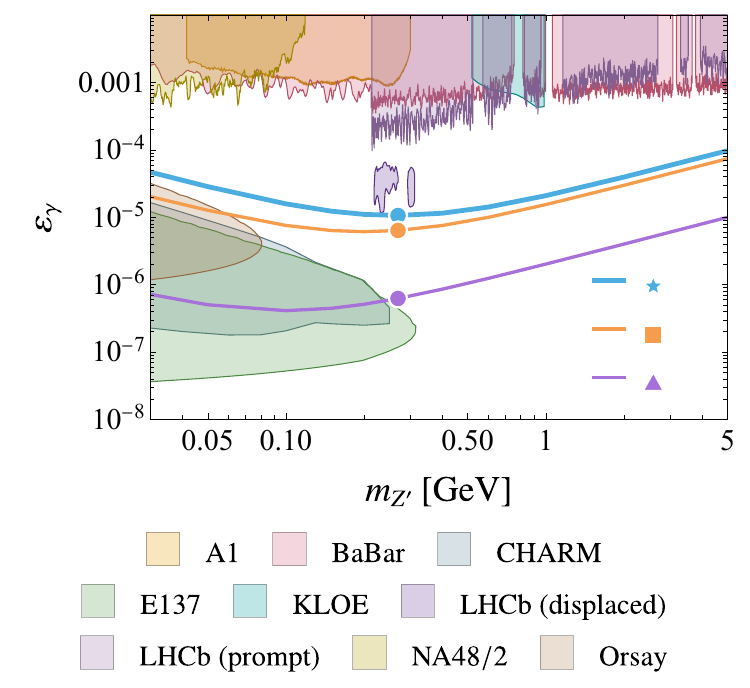}
  \caption{Signal-required $\varepsilon_\gamma$ for
  $\mathrm{BP}_\star$ (blue), $\mathrm{BP}_\square$ (orange) and
  $\mathrm{BP}_\triangle$ (purple), refitting the relic abundance
  and recoil normalization at each $m_{Z'}$.
  Shaded regions are the labelled visible dark-photon exclusions;
  the circles mark $m_{Z'}^{}=0.27~\mathrm{GeV}$ at
  the three benchmarks.}
  \label{fig:signal-zprime}
\end{figure}

The fitted recoil normalization also determines the kinetic mixing,
connecting these spectra to independent mediator searches.
\Cref{fig:signal-zprime} translates each benchmark recoil spectrum into a
required kinetic mixing $\varepsilon_\gamma^{}$ as the mediator mass
$m_{Z'}^{}$ varies. At each point, we iterate the relic solution and
fitted mixing to $1\%$ relative agreement. The curves therefore include
the change in the relic-fixed gauge coupling as well as the recoil
propagator. The colored circles mark the benchmark values at $m_{Z'}^{} =
0.27~\mathrm{GeV}$ listed in \cref{tab:benchmarks}. With $Z'\to a s_-$
closed, we compare the required mixing with visible dark-photon exclusions
\cite{Merkel:2014avp, BaBar:2014zli, KLOE-2:2018kqf, NA482:2015wmo,
LHCb:2019vmc, Tsai:2019buq, Andreas:2012mt, Davier:1989wz}, using DarkCast
tables \cite{Ilten:2018crw, Caputo:2026pdw}. The benchmarks evade the
exclusions shown.

\section{Conclusion}
\label{ipNGB:sec:conclusion}

We have studied a two-component dark sector containing a stable pNGB $a$
and its nearly degenerate quasi-pNGB partner $s_-$. The approximate scalar
symmetry connects the small mass splitting to the suppression of
tree-level Higgs-mediated elastic scattering. The coupled thermal
calculation yields comparable relic populations, while efficient
annihilation into light dark gauge boson $Z'$ reproduces the total
abundance. At $\mathrm{BP}_\star$, we have a substantial surviving excited
population under the adopted evolution assumptions. These ingredients
provide a thermal source of exothermic nuclear recoils with a suppressed
tree-level elastic signal.

Applied to the high-energy LZ event, the setup considered here admits a
relic-compatible benchmark whose exothermic recoil spectrum provides
support at the observed event energy. At $\mathrm{BP}_\star$, the expected
sideband yield is comparable to the search-window yield providing a
further test through future high-energy recoil measurements. Independent
dark gauge boson searches can also test the mixing required by this
interpretation.

\section*{Acknowledgements}
This work was supported in part by JSPS Grant-in-Aid for Scientific
Research KAKENHI Grants No. JP22K03620 (K.T.), 25H02179 (T.T.) and
25K07279 (T.T.). This work was supported by the Sasakawa Scientific
Research Grant from The Japan Science Society (R.S.).

\bibliographystyle{elsarticle-num}
\bibliography{references}

\end{document}